# Incomplete normalization of probability on multifractals

Q.A. Wang[1], L. Nivanen[1], A. El Kaabouchi[1], J.P. Badiali[2], A. Le Méhauté[1]

[1]Institut Supérieur des Matériaux et Mécaniques Avancées du Mans, 44 Av. Bartholdi, 72000 Le Mans, France

[2]UMR 7575 LECA ENSCP-UPMC,11 rue P. et M. Curie, 75231 Cedex 05, Paris, France

**Abstract**

This work is an extension of the incomplete probability theory from the simple case of monofractals previously studied to the more general case of multifractals which can occur in the phase space without equiprobable partition.

## 1) Introduction

The incomplete probability distribution (IPD) has been proposed for the case where the probability cannot be normalized, i.e., $\sum_{i=1}^{w} p_i \neq 1$. For a class of IPD, we have suggested the following incomplete normalization (IN)[1]:

$$\sum_{i=1}^{w} p_i^q = 1 \qquad (1)$$

The reader can find discussion of the various reasons for the existence of this nonadditive or unnormalizable probability distribution in the references [1][3][4][5][6][7].

One of the cases for which IN is suggested is the chaotic systems evolving in phase space having fractal attractors [2][3][4]. In our previous papers[3][4], IN was studied on monofractals having homogeneous distribution of segments of same length such as the standard Cantor set in Figure 1. This set is constructed by iteration. At the $k^{th}$ iteration, the initial line of length $L_0$ is transformed into a set with $N_k = 2^k$ segments of length $\delta_{ki} = \delta_k = \left(\frac{1}{3}\right)^k L_0$ ($i = 1, 2, \ldots, N_k$ are in the order say from left to the right). To understand the link between this structure evolution and the probability distribution of nonequilibrium systems, we can consider an ensemble of chaotic systems which evolve in their phase space[8] and are gradually attracted in a fractal structure (strange attractor) formed by their trajectories. At the same time, the occupied phase volume gradually maps into these attractors. To illustrate this evolution, let us take the example of a phase attractor in the form of the Cantor set of Figure 1. Suppose $L_0$ is the volume of the initial states, the total occupied volume at a later time $t_k$ of the $k^{th}$ iteration is $L_k = N_k \delta_k = \left(\frac{2}{3}\right)^k L_0$. If all the state points are equally probable, the usual definition of probability that a segment of length $\delta_{ki}$ is visited by the system is given by $P_{ki} = \frac{\delta_{ki}}{L_k}$. This probability can be normalized as follows :

$$\sum_{i=1}^{N_k} P_{ki} = \frac{\sum_{i=1}^{N_k} \delta_{ki}}{L_k} = \frac{N_k \delta_k}{L_k} = 1 \qquad (2)$$

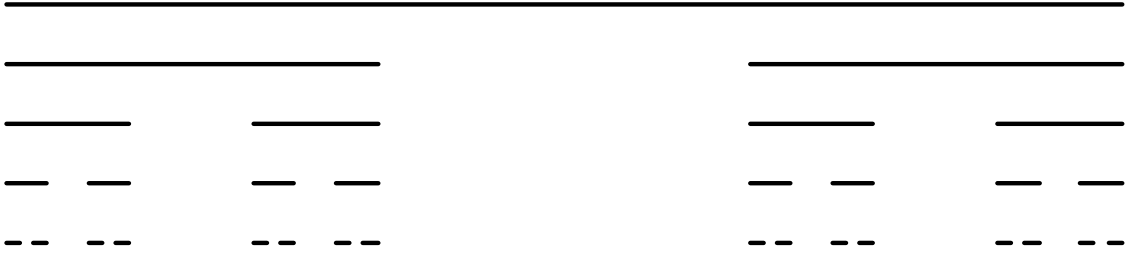

**Figure 1,** Standard Cantor set ($m = 2$, $\rho = 1/3$) built at iteration orders from $k = 0$ to $k = 4$. At first iteration, the total length of the curve is $L_1 = \left(\frac{2}{3}\right) L_0$ with $N_1 = 2$ segments of length $\delta_1 = \left(\frac{1}{3}\right) L_0$ ; at second iteration, $L_2 = \left(\frac{2}{3}\right)^2 L_0$ with $N_2 = 2^2$ segments of length $\delta_2 = \left(\frac{1}{3}\right)^2 L_0$, etc.

The probability defined in Eq.(2) is scientifically sound and in agreement with the probability theory. Its only drawback is that it is not convenient for describing physical systems out of equilibrium. If the system is in equilibrium at a stage $k$, the above definition of probability is reasonable and sufficient for the statistical description of the equilibrium state on the total phase volume $L_k$. However, if the system is out of equilibrium and $L_k$ is only an intermediate phase volume changing in time, it will be more convenient, in order to take into account the time evolution of the distribution, to define the probability finding a system in $\delta_{ki}$ with respect to the initial volume $L_0$[1][3] , i.e.,

$$p_{ki} = \frac{\delta_{ki}}{L_0}. \qquad (3)$$

This probability not only gives us the nonequilibrium distribution at any moment but also the evolution of the distribution with respect to the initial conditions. Notice that this probability is not normalizable since we have $\sum_{i=1}^{N_k} p_{ki} = \frac{\sum_{i=1}^{N_k} \delta_{ki}}{L_0} = \frac{L_k}{L_0} \neq 1$.

In our previous work[3], it was shown that, for the case of self-similar monofractal, the probability of Eq.(3) can be normalized by :

$$\sum_{i=1}^{N_k} p_{ki}^{\,q} = 1 \qquad (4)$$

for any stage $k$ with a unique $q = d$, where $d$ is the self-similarity dimension of the monofractals (see definition below). The parameter $q$ characterizes the evolution of the phase space. When $q < 1$, the phase volume shrinks in time, when $q > 1$, there is phase space expansion. The normalization becomes complete for $q = 1$ without evolution of distribution. Eq.(4) has been derived on monofractals having segments of same length such as in Figure 1. It was only conjectured[3] that the same calculation would hold for multifractals which in general contain segments of different lengths, but no proof has been given.

In this work, it will be shown that Eq.(4) holds for any multifractal with self-similar structure having the same iteration rule at each stage. In order to be clear for the readers who have not followed the previous work, we will begin by the simplest case of homogeneous monofractals. More complicated cases with different segments will be analyzed first with two segments of different lengths replacing a segment of precedent stage, and then with an arbitrary number $m$ of different segments. Proof for the uniqueness of $q$ will also be given.

## 2) Monofractals

In this section, we recapitulate some essential results obtained previously for the case of monofractals. Let us consider a deterministic mass fractal form $F$ such as the usual Cantor set on Figure 1. Let $F_k$ be the intermediate $k^{th}$ stage of construction. When the initiator $F_0$ is a segment of length $L_0$, $F_1$ is defined by $N_1 = m$ copies of $F_0$ with same scale factor $\rho$. We suppose that two different copies never intersect. At the $k^{th}$ stage, $F_k$ is composed of $N_k = m^k$ segments of same length $\delta_{ki} = \delta_k = \rho^k L_0$. The total length of the $k^{th}$ stage $F_k$ form is given by :

$$L_k = \sum_{i=1}^{N_k} \delta_{ki} = N_k \delta_k = (m\rho)^k L_0 \qquad (5)$$

The notion of self-similar dimension is defined from the idea of fractal content of order $\alpha$ of the form $F_k$ :

$$C_{k,\alpha} = \sum_{i=1}^{N_k} \delta_{ki}{}^\alpha = N_k \delta_k{}^\alpha = m^k (\rho^k L_0)^\alpha = \left(m\rho^\alpha\right)^k L_0{}^\alpha \tag{6}$$

The fractal similarity dimension $d$ is derived from the behavior of the content in the $k \to +\infty$ limit :

$$\begin{cases} \alpha > d \; : C_{k,\alpha} \to 0 \\ \alpha < d \; : C_{k,\alpha} \to +\infty \end{cases} \tag{7}$$

In the monofractal case, $m\rho^d = 1$ so that $d = \dfrac{\ln m}{\ln(1/\rho)}$.

Now let us consider two approaches leading to two different measures.

First we only look at the stage $k$ and choose a point $M$ on $F_k$. As $F_k$ is composed of $N_k$ segments, $M$ belongs to one of these segments. As mentioned above and used in many textbooks[8], the measure $P_{ki} = \dfrac{\delta_{ki}}{L_k}$ is a probability. The monofractal character of $F$ implies that at any stage $P_{ki} = \dfrac{\delta_k}{L_k} = P_k$ is independent of $i$. It is given by :

$$P_{ki} = P_k = \frac{\delta_{ki}}{L_k} = \frac{\rho^k L_0}{(m\rho)^k L_0} = \frac{1}{m^k} \tag{8}$$

and normalized by :

$$\sum_{i=1}^{N_k} P_{ki} = N_k P_k = m^k \frac{1}{m^k} = 1. \tag{9}$$

This distribution $\{P_{ki}\}$, $1 \leq i \leq N_k$, only concerns the $k^{th}$ stage $F_k$ of the monofractal form. The fractal character of the distribution and of its time evolution is not taken into account.

Now let us look at the fractal building in an evolutionary point of view by choosing a point $M$ on the initiator $F_0$ and following its behavior in time. In a mass fractal like Cantor set, we have $F_k \subset F_{k-1}$ for any $k$, which implies $F_k \subset F_0$. So $M$ may or may not belong to the $k^{th}$ stage $F_k$. This is a kind of loss of state points (or information). To evaluate this evolution, we introduce the quantity $p_{ki} = \dfrac{\delta_{ki}}{L_0}$ [3]. The monofractal character of $F$ implies that at any stage $p_{ki} = \dfrac{\delta_k}{L_0} = p_k$ is independent of *i*. We get from the definition of fractal dimension :

$$p_{ki}{}^d = \left(\frac{\delta_k}{L_0}\right)^d = \left(\frac{\rho^k L_0}{L_0}\right)^d = \rho^{kd} = \frac{1}{m^k} \tag{10}$$

and

$$\sum_{i=1}^{N_k} p_{ki}{}^d = N_k p_k{}^d = m^k \frac{1}{m^k} = 1 \tag{11}$$

The distribution $\{p_{ki}\}$, $1 \leq i \leq N_k$ is essentially different from the normalizable probability distribution Eq.(8). As indicated above, it measures the probability to encounter the $i^{th}$ segment of $F_k$ when the system comes from somewhere on the initial set $F_0$. It takes into account the fractal nature of the evolution of set $F$.

## 3) Binary multifractals

As in the case of monofractal sets, a probability law with IN can be defined on a multifractal set characterized by different scale factors for each copy of the initial segment. In this section, we first treat the case of a "binary" multifractal set. At each stage, a segment is replaced by $m = 2$ copies with scale factors $\rho_1$ and $\rho_2$, respectively ($0 < \rho_1, \rho_2 < 1$). An example is given on Figure 2.

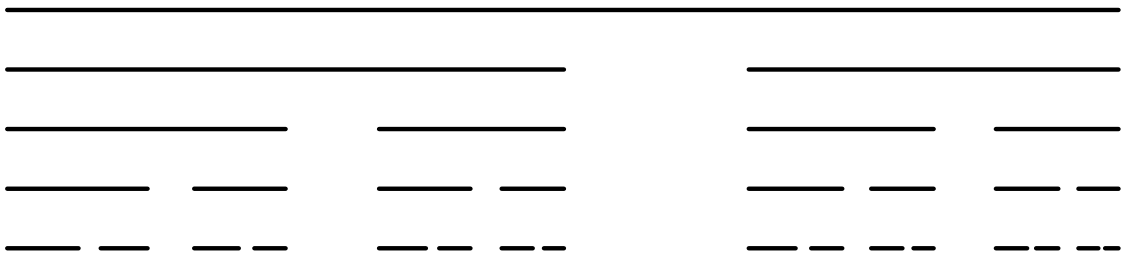

**Figure 2,** Modified binary multifractal Cantor set ($m = 2$, $\rho_1 = 1/2$, $\rho_2 = 1/3$) built at iteration orders from $k = 0$ to $k = 4$.

At the $k^{th}$ stage, $F_k$ is composed of $N_k = 2^k$ segments. Let $j \in \{0,\ldots, k\}$. On $F_k$, we can find $\binom{k}{j}$ segments of length $\Delta_{k,j} = \rho_1^{k-j} \rho_2^{j} L_0$. The total length of the $k^{th}$ stage $F_k$ form is given by:

$$L_k = \sum_{i=1}^{N_k} \delta_{ki} = \sum_{j=0}^{k} \binom{k}{j} \Delta_{k,j} = (\rho_1 + \rho_2)^k L_0 \tag{12}$$

The notion of self-similarity dimension can be extended to the case of a multifractal set. The content of order $\alpha$ of the form $F_k$ takes the form:

$$C_{k,\alpha} = \sum_{i=1}^{N_k} \delta_{ki}^{\alpha} = \sum_{j=0}^{k} \binom{k}{j} \Delta_{k,j}^{\alpha} = \sum_{j=0}^{k} \binom{k}{j} \rho_1^{(k-j)\alpha} \rho_2^{j\alpha} L_0^{\alpha} \tag{13}$$

$$C_{k,\alpha} = \sum_{j=0}^{k} \binom{k}{j} (\rho_1^{\alpha})^{k-j} (\rho_2^{\alpha})^{j} L_0^{\alpha} = (\rho_1^{\alpha} + \rho_2^{\alpha})^k L_0^{\alpha} \tag{14}$$

The multifractal similarity dimension $d$ is introduced by Eq.(7). It verifies:

$$\rho_1^{d} + \rho_2^{d} = 1 \tag{15}$$

The quantities $P_{ki}$ and $p_{ki}$ are introduced in the same way as in Section 2, by setting $P_{ki} = \dfrac{\delta_{ki}}{L_k}$ and $p_{ki} = \dfrac{\delta_{ki}}{L_0}$ respectively. We note that the distributions $\{P_{ki}\}$ and $\{p_{ki}\}$, $1 \leq i \leq N_k$ are no longer homogeneous due to the different scale factors $\rho_1$ and $\rho_2$. Indeed, they verify the same normalization relationships as Eqs.(9) and (11) in Section 2. Again, $P_{ki}$ is normalizable:

$$\sum_{i=1}^{N_k} P_{ki} = \frac{1}{L_k} \sum_{i=1}^{N_k} \delta_{ki} = \frac{1}{L_k} L_k = 1 \tag{16}$$

For the calculation of the sum of $p_{ki}^{d}$, let us classify all segments (denoted by $i$) of $F_k$ into groups (denoted by $j$) of segments of same length $\Delta_{k,j}$, such that the sum over $i$ be replaced by a sum over $j$, i.e.,

$$\sum_{i=1}^{N_k} p_{ki}{}^d = \frac{1}{L_0{}^d}\sum_{i=1}^{N_k}\delta_{ki}{}^d = \frac{1}{L_0{}^d}\sum_{j=0}^{k}\binom{k}{j}\Delta_{k,j}{}^d = \frac{1}{L_0{}^d}\sum_{j=0}^{k}\binom{k}{j}\rho_1{}^{(k-j)d}\rho_2{}^{jd}L_0{}^d \qquad (17)$$

or

$$\sum_{i=1}^{N_k} p_{ki}{}^d = \sum_{j=0}^{k}\binom{k}{j}\left(\rho_1{}^d\right)^{k-j}\left(\rho_2{}^d\right)^{j} = \left(\rho_1{}^d + \rho_2{}^d\right)^k = 1^k = 1. \qquad (18)$$

So in the binary multifractal case, a probability distribution with incomplete normalization such as Eq.(1) can still be defined according to Eq.(18) at each stage of a given multifractal with $q=d$.

Since the factor $(\rho_1 + \rho_2)^k$ appears in Eq.(12), all the distributions $\{P_{ki}\}$, $1 \leq i \leq N_k$ can be derived from the distribution of $F_1$ stage $\{P_{1i}\}$, $1 \leq i \leq 2 = N_1$. For instance, from Figure 2, we get $P_{11} = 3/5$, $P_{12} = 2/5$ and $P_{21} = 9/25 = P_{11}{}^2$, $P_{22} = P_{23} = 6/25 = P_{11}P_{12}$, $P_{24} = 4/25 = P_{12}{}^2$, respectively. In the same way, the factor $\left(\rho_1{}^d + \rho_2{}^d\right)^k$ appears in Eq.(18), so all the distributions $\{p_{ki}\}$, $1 \leq i \leq N_k$ can be derived from the $F_1$ stage distribution $\{p_{1i}\}$, $1 \leq i \leq 2 = N_1$. From Figure 2, we have $p_{11} = 3/5$, $p_{12} = 2/5$, and $p_{21} = 9/25 = p_{11}{}^2$, $p_{22} = p_{23} = 6/25 = p_{11}p_{12}$, $p_{24} = 4/25 = p_{12}{}^2$. It means that for a given fractal $F$, the essential relationships for Eqs.(16) and (18) to hold are the iteration rules applied to $k = 1$ and to each following stage of the structure.

## 4) General multifractals

In this section, we consider the case of general deterministic multifractal sets. At each stage, each segment is replaced by $m$ copies with scale factors $\rho_1, \rho_2, ..., \rho_m$ respectively. For any integer $i \in \{1, 2, ..., m\}$, $0 < \rho_i < 1$. Such an example is given on Figure 3.

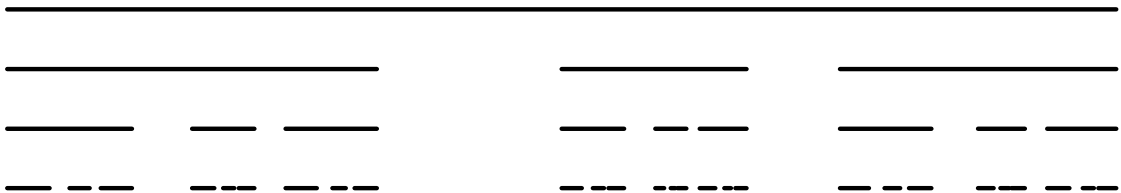

**Figure 3,** Modified multifractal Cantor set ($m = 3$, $\rho_1 = 1/3$, $\rho_2 = 1/6$, $\rho_3 = 1/4$) built at iteration orders from $k = 0$ to $k = 3$.

At the $k^{th}$ stage, $F_k$ is composed of $N_k = m^k$ segments of length $\delta_{ki}$. We note $J$ any set of integers

$\{j_1, j_2, ..., j_m\}$, that verify :

$$\forall i \in \{1, 2, ..., m\}, j_i \in \{0, 1, ..., k\}, \quad \sum_{i=1}^{m} j_i = k \quad (19)$$

For any set $J$ that verifies Eq.(19), we can find $\dfrac{k!}{\prod_{i=1}^{m} j_i!}$ segments of length $\Delta_{k,J} = \left(\prod_{i=1}^{m} \rho_i^{j_i}\right) L_0$ on $F_k$. The total length of the $F_k$ form at $k^{th}$ stage is given by :

$$L_k = \sum_{i=1}^{N_k} \delta_{ki} = \sum_J \left(\dfrac{k!}{\prod_{i=1}^{m} j_i!}\right)\left(\prod_{i=1}^{m} \rho_i^{j_i}\right) L_0 \quad (20)$$

where the sum is extended upon the all possible sets of the form in Eq.(19). We see a multinomial series expansion such as

$$L_k = \left(\sum_{i=1}^{m} \rho_i\right)^k L_0 \quad (21)$$

The content of order $\alpha$ of the form $F_k$ can be performed :

$$C_{k,\alpha} = \sum_{i=1}^{N_k} \delta_{ki}^{\alpha} = \sum_J \left(\dfrac{k!}{\prod_{i=1}^{m} j_i!}\right)\left(\prod_{i=1}^{m} \rho_i^{j_i}\right)^{\alpha} L_0^{\alpha} = \sum_J \left(\dfrac{k!}{\prod_{i=1}^{m} j_i!}\right)\left(\prod_{i=1}^{m} (\rho_i^{\alpha})^{j_i}\right) L_0^{\alpha} = \left(\sum_{i=1}^{m} \rho_i^{\alpha}\right)^k L_0^{\alpha} \quad (22)$$

We deduce the identity verified by the multifractal similarity dimension $d$ :

$$\sum_{i=1}^{m} \rho_i^{d} = 1 \quad (23)$$

In order to see the uniqueness of $d$, let us define $f$ by $f(\alpha) = \sum_{i=1}^{m} \rho_i^{\alpha}$ which is a strictly decreasing and continuous function. We have in addition $f(\alpha) = m$ for $\alpha = 0$ and $f(\alpha) \to 0$ for $\alpha \to +\infty$, meaning that Eq.(23) is satisfied by a unique real $d$.

The quantities $P_{ki}$ and $p_{ki}$ are defined in the same way as in Sections 2) and 3), by setting $P_{ki} = \dfrac{\delta_{ki}}{L_k}$ and $p_{ki} = \dfrac{\delta_{ki}}{L_0}$ respectively. The incomplete normalization relationship Eq.(18) can be generalized into :

$$\sum_{i=1}^{N_k} p_{ki}{}^d = \dfrac{\sum_{i=1}^{N_k}\delta_{ki}{}^d}{L_0{}^d} = \dfrac{1}{L_0{}^d}\sum_J \left(\dfrac{k!}{\prod_{i=1}^{m} j_i!}\right)\Delta_{k,J}{}^d = \dfrac{1}{L_0{}^d}\sum_J \left(\dfrac{k!}{\prod_{i=1}^{m} j_i!}\right)\left(\prod_{i=1}^{m}\rho_i{}^{j_i}\right)^d L_0{}^d \qquad (24)$$

$$= \sum_J \left(\dfrac{k!}{\prod_{i=1}^{m} j_i!}\right)\left(\prod_{i=1}^{m}(\rho_i{}^d)^{j_i}\right)$$

which is a multinomial series expansion :

$$\sum_{i=1}^{N_k} p_{ki}{}^d = \left(\sum_{i=1}^{m}\rho_i{}^d\right)^k = 1^k = 1 \qquad (25)$$

As in the binary case, an IPD with a unique $q$ can be defined on a general deterministic multifractal set with the same rule of iteration at each stage. For all the cases considered up to now, the uniqueness of $q$ comes from the uniqueness of the fractal dimension $d$. The multifractal case is particularly interesting because it makes it possible to extend incomplete distribution beyond the simple case of equiprobable phase space partition.

## 5) Multifractals with multidimensional initiators

The above cases concern only one dimensional initiator $F_0$. In this section, we address the multifractals having multidimensional initiators. The multidimensional monofractals have been studied in a previous work[3], which we will recapitulate in what follows before applying it to the multifractal case. An example of multidimensional monofractal defined by a unique scale factor $\rho$ from a square initiator is given in Figure 4.

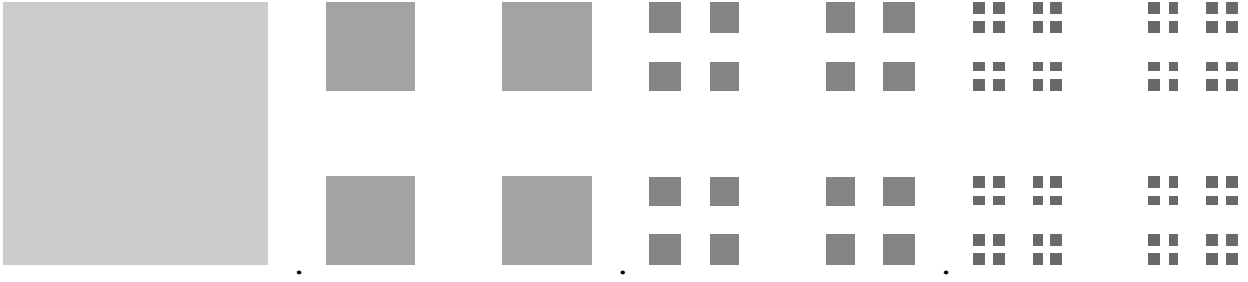

**Figure 4,** bidimensional Cantor set ($m = 2$, $\rho = 1/3$) built at iteration orders from $k = 0$ to $k = 3$.

The fractal dimension is still introduced from a length measure. In Figure 4, the length $L_{ki} = L_k = \rho^k L_0$ is given by the edge of any square of the fractal form $F_k$. For a more general initiator, we choose a characteristic length and call it a gauge. The fractal dimension is still defined by Eqs.(6) and (7).

In the plane, $F_0$ is a polygon of surface $S_0$ and $F_k$ is composed of $N_k = m^k$ polygons of surface $\sigma_{ki} = \sigma_k = \rho^{2k} S_0$. The total surface $S_k$ of the $k^{th}$ stage is given by :

$$S_k = \sum_{i=1}^{N_k} \sigma_{ki} = N_k \sigma_k = \left(m\rho^2\right)^k S_0 \qquad (26)$$

The quantities $P_{ki}$ and $p_{ki}$ are now defined with respect to the surfaces, i.e., $P_{ki} = \dfrac{\sigma_{ki}}{S_k}$ and $p_{ki} = \dfrac{\sigma_{ki}}{S_0}$ respectively.

The distribution $\{p_{ki}\}$ still verifies an incomplete normalization relationship with a new exponent $d/2$. Indeed

$$p_{ki}{}^{d/2} = \left(\frac{\rho^{2k} S_0}{S_0}\right)^{d/2} = \rho^{kd} = \frac{1}{m^k} \Rightarrow \sum_{i=1}^{N_k} p_{ki}{}^{d/2} = 1 \qquad (27)$$

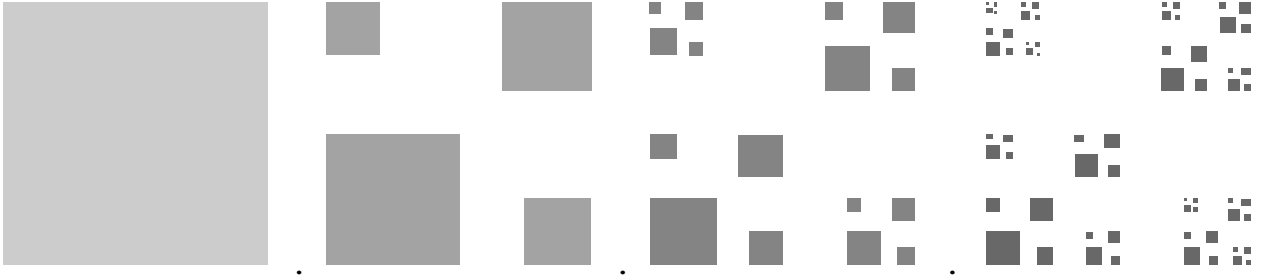

**Figure 5,** bidimensional multifractal Cantor set ($m = 4$, $\rho_1 = 1/5$, $\rho_2 = 1/3$, $\rho_3 = 1/2$, $\rho_4 = 1/4$) built by iteration orders from $k = 0$ to $k = 3$.

Now let us consider a bidimensional multifractal set like on Figure 5. The fractal dimension is still defined from a gauge, and verifies Eq.(23). The quantities $P_{ki}$ and $p_{ki}$ are defined in the same way as for the bidimensional monofractal case, by $P_{ki} = \dfrac{\sigma_{ki}}{S_k}$ and $p_{ki} = \dfrac{\sigma_{ki}}{S_0}$ respectively.

Let $S_0$ be the surface of the initiator $F_0$. At the $k^{th}$ iteration, the form $F_k$ is composed of $N_k = m^k$ copies of $F_0$ of surface $\sigma_{ki}$. For any set $J$ that verifies Eq.(19), we can find $\dfrac{k!}{\prod_{i=1}^{m} j_i!}$ copies of $F_0$ of surface $S_{k,J} = \left( \prod_{i=1}^{m} \rho_i^{2 j_i} \right) S_0$ on $F_k$. The total surface of the $k^{th}$ stage $F_k$ form is given by :

$$S_k = \sum_{i=1}^{N_k} \sigma_{ki} = \sum_{J} \left( \dfrac{k!}{\prod_{i=1}^{m} j_i!} \right) \left( \prod_{i=1}^{m} \rho_i^{2 j_i} \right) S_0 \qquad (28)$$

where the sum is carried out over all the possible sets of the form Eq.(19). We get a multinomial series expansion :

$$S_k = \left( \sum_{i=1}^{m} \rho_i^2 \right)^k S_0. \qquad (29)$$

The incomplete normalization relationship Eq.(25) now becomes :

$$\sum_{i=1}^{N_k} p_{ki}^{d/2} = \frac{\sum_{i=1}^{N_k} \sigma_{ki}^{d/2}}{S_0^{d/2}} = \frac{1}{S_0^{d/2}} \sum_J \left( \frac{k!}{\prod_{i=1}^{m} j_i!} \right) S_{k,J}^{d/2} \qquad (30)$$

i.e.,

$$\sum_{i=1}^{N_k} p_{ki}^{d/2} = \frac{1}{S_0^{d/2}} \sum_J \left( \frac{k!}{\prod_{i=1}^{m} j_i!} \right) \left( \prod_{i=1}^{m} \rho_i^{2 j_i} \right)^{d/2} S_0^{d/2} = \sum_J \left( \frac{k!}{\prod_{i=1}^{m} j_i!} \right) \left( \prod_{i=1}^{m} (\rho_i^d)^{j_i} \right) \qquad (31)$$

which is a multinomial series expansion :

$$\sum_{i=1}^{N_k} p_{ki}^{d/2} = \left( \sum_{i=1}^{m} \rho_i^d \right)^k = 1^k = 1 \qquad (32)$$

An incomplete probability distribution with unique $q=d/2$ (still resulting from the uniqueness of the fractal dimension) can be defined on a general bidimensional multifractal set with the same rule of iteration at each stage. This result can be extended to higher dimensional initiators. For the "three dimensional" case for example, one obtains an exponent $q=d/3$. For arbitrary topological dimension $d_T$, we should have :

$$\sum_{i=1}^{N_k} p_{ki}^{d/d_T} = 1 \qquad (33)$$

## 6) Concluding remarks

The present work is an extension of the IPD from the simple case of monofractals to the multifractals. This extension is useful for the understanding of the incomplete statistics and of their relevance in the study of nonequilibrium complex systems evolving in fractal geometry such as the strange attractors in phase space. The conclusion of the present work is that the mathematical formalism of IPD having incomplete normalization with a unique parameter $q=d/d_T$ is valid for any deterministic multifractal with the same rule at each iteration.

However, the uniqueness of $q$ may be perturbed in the case of the fractals which change rules during their iterations or for different scales. The possible relationships between $q$ and the different iteration rules and the consequences of this multi-rule evolution are under investigation. The result is expected to be helpful for the understanding of the evolution of many complex systems with obvious dynamical transition in time and in scale, such as the degree distributions in multiple regimes observed in many scale free networks.